\documentclass[twocolumn,preprintnumbers,
endnote,nofootinbib,prl]{revtex4}
\usepackage{graphicx}

\usepackage[hypertex]{hyperref}
\newcommand{\vev}[1]{\langle {#1} \rangle}
\newcommand{\lsim}{\lesssim}
\newcommand{\gsim}{\gtrsim}

\newcommand{\eq}[1]{Eq.~(\ref{#1})}

\newcommand{\beq}{\begin{equation}}
\newcommand{\eeq}{\end{equation}}
\newcommand{\eps}{\varepsilon}

\newcommand{\rmP}{\bar{M}_{\rm Pl}}

\begin{document}

\pagestyle{plain}

\title{Gravitationally Induced Dark Matter Asymmetry and Dark Nucleon Decay}

\author{Hooman Davoudiasl\footnote{email: hooman@bnl.gov}
}
\affiliation{Department of Physics, Brookhaven National Laboratory,
Upton, NY 11973, USA}


\begin{abstract}

The ``gravitational baryogenesis" scenario is extended to generate both baryon and dark matter
asymmetries, in the matter dominated era corresponding to post-inflationary reheating.
A minimal extension requires a singlet fermion $X$ for dark matter and a singlet scalar $S$.
With two or more hidden sector fermions, the scenario can lead to
{\it nucleon decay into dark matter} with a lifetime of order $10^{34-36}$~yr,
which is relevant for current or future experiments.
The correct multi-component relic density can be obtained if dark matter fermions couple to a sub-GeV vector
boson that weakly interacts with the Standard Model through mixing.  The typical inflationary scale in the scenario
is of order $10^{16}$~GeV which suggests that tensor mode perturbations could potentially be within observational reach.

\end{abstract}
\maketitle

The cosmological energy densities
of baryons and dark matter (DM), respectively denoted by $\Omega_B$ and $\Omega_{DM}$, have
similar sizes \cite{PDG,Planck}
\beq
\frac{\Omega_{DM}}{\Omega_B} \approx 5\,,
\label{ODMOB}
\eeq
even though they have very different properties.
This empirical fact provides motivation for postulating a common
origin for cosmic baryon and dark matter
abundances.  A common origin suggests that the DM number density $n_{DM}$, like the number density
$n_B$ of baryons, is given by an asymmetry.
Various cosmological observations imply that the ratio of baryon
density to that of entropy $s$ is given by \cite{PDG}
\beq
\frac{n_B}{s} \approx 10^{-10}\,.
\label{nB/s}
\eeq
If the origin of the baryon and DM asymmetries
is the same, we may expect $n_B\sim n_{DM}$ which implies 
an asymmetric dark matter (ADM) \cite{ADM} mass
in the GeV regime.  A variety of mechanisms for ADM have been proposed in the literature.  See
Ref.~\cite{ADM_early} for some of the pioneering work in this direction and 
Refs.~\cite{ADM_recent, ADM,Shelton:2010ta, hylogenesis} for a sample of more recent investigations.  
Ref.~\cite{reviews} contains some reviews of the subject.  

In what follows, we will consider extending the ``gravitational baryogenesis"
mechanism proposed in  Ref.~\cite{GB} to include the generation of
a DM asymmetry.  In this scenario, dynamical violation of CPT in an
expanding universe leads to the generation of
asymmetries, in thermal equilibrium, through the coupling \cite{GB}
\beq
\frac{1}{M_c^2} \int d^4 x \sqrt{-g}
(\partial_{\mu}{\cal R})J_Q^{\mu},
\label{OQ}
\eeq
where $M_c$ is the gravity cutoff scale,
${\cal R}$ is the Ricci scalar curvature, and
$J_Q^\mu$ is the current associated with a quantum number $Q$.  The scale $M_c$
is typically of order the reduced Planck mass
$\rmP \approx 2.4\times 10^{18}$~GeV, but could be somewhat different.
The universality of gravitational interactions suggests
that such couplings generally exist.

We will assume that $Q=B$ is the baryon number in the visible sector.
Let $X$ be a Dirac fermion, assumed to be DM, carrying a hidden charge $X=+1$.

Following the arguments presented in Ref.~\cite{GB},
the interaction in Eq.~(\ref{OQ}) provides a bias in thermal
equilibrium that acts as a chemical potential for generating a charge asymmetry.  This
is due to dynamical CPT violation from the time evolution
of the Ricci scalar (this is similar in spirit to ``spontaneous baryogenesis" \cite{Cohen:1987vi}).
The baryon asymmetry is then given by \cite{GB}
\beq
\frac{n_B}{s} \approx \left.
\frac{\dot{{\cal R}}}{M_c^2 T}\right|_{T_D},
\label{nB}
\eeq
for $T<T_D$, where $T_D$ is the temperature at which processes that violate $B$
decouple; a dot represents a time derivative.  For
$B$ violation, we will consider the dim-6 operator \cite{xudd}
\beq
O_{BX} = \frac{(Xudd)_R}{\Lambda^2} + \small{\rm H.C.}\,,
\label{OBX}
\eeq
where $u$ and $d$ are the up- and down- type quarks in the Standard Model (SM), the subscript $R$
denotes right-handed chirality, and generation and color indices have been
suppressed; $\Lambda$ is the scale of $B+X$ violation.
Note that $O_{BX}$ preserves $B-X$, which
we will assume to be a good symmetry.  This assumption excludes neutrino Dirac mass
operators of the type $H L X$, where $H$ is the Higgs field and $L$ is a lepton doublet of the SM.

The induced CPT violation from Eq.~(\ref{OQ}) and the interaction in Eq.~(\ref{OBX}) lead to the generation of equal
asymmetries in $X$ and $B$.  The asymmetry in $X$ remains unprocessed.  However, the $B$ number
can get partially converted into lepton number if the electroweak sphaleron processes are active, corresponding
to a reheat temperature $T_{\rm R} \gsim 100$~GeV, after inflation.
For $T_{\rm R} \gsim 100$~GeV, the well-known results of Ref.~\cite{Harvey:1990qw} then yield
\beq
n_B = (28/79) \, n_X
\label{nBnX}
\eeq
for the size of the baryon asymmetry at $T\ll 100$~GeV.
Eq.~(\ref{ODMOB}) then implies that $m_X \approx 2$~GeV if
the energy density of DM is set by the value of its asymmetry.
On the other hand, if $T_{\rm R} \lsim 100$~GeV,
the sphalerons are out of thermal equilibrium and the relation
$n_B = n_X$ is maintained down to low temperatures,
implying that $m_X \approx 5$~GeV.

In order for the DM density to be set by the asymmetry $n_X$, we need the symmetric
population of $X$ and $\bar{X}$ particles to annihilate away.
As a first attempt, we simply assume that there is a singlet
scalar $S$ that couples to $X$ and gives it mass.  Let the
couplings of $X$, $S$, and the SM Higgs doublet $H$ be given by
\beq
{\cal L}=y_X \, S {\bar X} X\ + \lambda_S S^2 H^\dagger H\,.
\label{Lmass}
\eeq
If $m_X > m_S$ the annihilation process $X\bar X \to S S$ could be used
for depleting the symmetric $X$ population.  Since we are interested in
$m_S \lsim 1$~GeV, we require $\lambda_S \lsim 10^{-5}$ in order to
avoid tuned cancelations in the potential for $S$ with
$\vev{H}\approx 246$~GeV in \eq{Lmass}.  However, $S$ needs to decay well
before Hubble time $t_H\sim 1$~s, the onset of Big Bang Nucleosynthesis,
to avoid large deviations from standard cosmology.

The mass mixing parameter
\beq
\mu^2 = \lambda_S \vev{H}\vev{S}
\label{mu2}
\eeq
from the second term in Eq.~(\ref{Lmass})
leads to the mixing of the Higgs $H$ with the singlet $S$, given by the angle
\beq
\xi \sim \frac{\mu^2}{\vev{H}^2}\,.
\label{xi}
\eeq
We will be interested in the case
$m_S\sim \vev{S}\lsim 1$~GeV which implies $\xi \lsim 10^{-7}$.
To estimate an upper bound on the lifetime $\tau_S$ of $S$, let us consider the
decay into $\mu^+\mu^-$, via mixing with the Higgs.  We then have
\beq
1/\Gamma(S\to \mu^+ \mu^-)\approx \frac{16 \pi}{\xi^2 y_\mu^2 m_S}
\gsim 10^{-2}~{\rm s}\,,
\label{GamStoss}
\eeq
where the muon
Yukawa coupling is given by $y_\mu \approx 5\times 10^{-4}$.
The above bound allows for sufficiently fast decay of $S$; $\tau_S\ll 1$~s.  However, below
we will also consider cases where $S$ may need to be somewhat
lighter than $\sim 1$~GeV for efficient $X\bar X$ annihilation.  In that case, without tuning
the $S$ mass parameter, $\xi$ would need to be somewhat smaller than
$10^{-7}$ which will lead to $\tau_S\gsim 1$~s.  We will address this question
near the end of this work,
by considering annihilation into dark $U(1)_d$ vector bosons that
kinetically mix with the photon \cite{Holdom:1985ag}.

The cross section for $X\bar X \to SS$ is given by
\beq
\sigma(X \bar X\to S S) \sim
\frac{y_X^4}{32 \pi m_X^2}\,.
\label{XtoS}
\eeq
For $m_X \approx 2$~GeV and $y_X \sim 1$, the above equation yields
$\sigma_{X\bar X} \sim ~10^{-3}{\rm GeV}^{-2} \sim $~$\mu$b.  We note that this
is much larger than a typical thermal relic annihilation cross section $\sim 1$~pb.  Hence, with
our typical assumptions, an efficient depletion of the symmetric DM population
can be expected in this minimal setup.  We will next consider
a cosmological context that could lead to sufficient dynamical CPT violation required
for the generation of the relic asymmetries.

According to Eq.~(\ref{nB}), the value of asymmetry achieved through
gravitational genesis depends on cosmological
evolution through  $\dot{\cal R}$, which is given by \cite{GB}
\beq
\dot{\cal R} = -(1-3w)\frac{\dot{\rho}}{\rmP^2} =
\sqrt{3}\, (1-3w)(1+w)\frac{\rho^{3/2}}{\rmP^3},
\label{Rdot}
\eeq
where $w$ is the ratio of pressure and energy density $\rho$.
As an illustrative example, let us consider $w=0$, corresponding
to matter domination.  Post-inflationary reheating during
inflaton oscillations is described by a matter dominated equation of state and
hence $w=0$ is a well motivated choice.
Using the result derived for this case in Ref.~\cite{GB} we get
\beq
\frac{n_B}{s} \approx \frac{T_D^6}{M_c^2 \rmP^3 T_{\rm R}},
\label{nBs;w=0}
\eeq
which is obtained in a linear approximation, valid for $T_{\rm R} \gsim 10^{-2} T_D$.  We recall that
$T_D$ is the temperature at which $B$ violation mediated by the 
interaction in Eq.~(\ref{OBX}) becomes decoupled.  
Assuming $M_c\sim \rmP$, one can then obtain
the requisite $n_B/s$ with $T_{\rm D} \sim 2\times 10^{16}$~GeV and $T_{\rm R}\sim 10^{16}$~GeV.
For these values, one gets $\Lambda\sim {\rm few}\times 10^{16}$~GeV.
Here, the energy density at $T=T_{\rm D}$ is given by $\rho \sim T_{\rm D}^8/T_{\rm R}^4$ \cite{GB}.  Hence,
these values of parameters are consistent with an inflationary scale $V_I^{1/4} \sim 10^{16}$~GeV.
Note that much larger inflationary scales may lead to excessive levels of tensor mode perturbations
\cite{Linde:1983gd,Lyth:1984yz,Planck}.

Since $T_{\rm R}\gg 100$~GeV, sphaleron processes will be active at $T<T_R$
after a baryon asymmetry has been generated.  Hence, baryon number
density is given by Eq.~(\ref{nBnX}), implying
that $m_X\approx 2$~GeV, for which
nucleon decay through the operator in Eq.~(\ref{OBX}) is not
relevant.  However,  the dark matter particle $X$ is unstable and can decay into a nucleon
and a meson with a very long
lifetime $\tau_X \sim 10^{34}$~yr.  The corresponding decay rate
of $X$ is well-below what can be detected through astrophysical observations or otherwise.
We also note that $S$-$H$ mixing, without tuning,
is typically suppressed by $\xi \lsim 10^{-7}$ and hence likely unobservable.
Thus, it seems that the minimal scenario of gravitationally generated asymmetric
dark matter $X$ is largely inaccessible to experimental tests.

The above conclusion can change in an interesting way if ADM is made up of
multiple fields.  Let us assume that there are $n$
dark fermions $\tilde X_i$, $i=1,2,\ldots,n$.
One could then generalize Eq.~(\ref{OBX}) to
\beq
O_{B \tilde X_i} = \sum_i^n \frac{(\tilde X_i udd)_R}{\Lambda_i^2} + \small{\rm H.C.}\,,
\label{OBXi}
\eeq
where $\Lambda_i$ are scales of $B+X$ violation corresponding to $\tilde X_i$.
If there is a hierarchy among $\Lambda_i$, then
Eq.~(\ref{nBs;w=0}) suggests that the $B$ violating interaction with the largest $\Lambda_i$ will dominate and
we are basically back to the minimal model.  However, if the ultraviolet model that
generates $O_{B \tilde X_i}$ is flavor symmetric, we then expect
$\Lambda_i$ to be of similar size.  For simplicity, let us
focus on a case in which there is such a symmetry for two fields $\tilde X_1$ and $\tilde X_2$.   In the
limit $\Lambda_1 \approx \Lambda_2$, we then get a two component ADM population with
$n_{\tilde X_1} \approx n_{\tilde X_2}$.

For $\vev{S}\neq 0$, Yukawa interactions of the type
$y_{ij} S \tilde X_i^c \tilde X_j$
yield a mass matrix with diagonal entries $m_1$ and $m_2$ and the
off diagonal entries $m_{12}$ and $m_{21}$.
We will refer to the mass eigenstates as $X_1$ and $X_2$,
with masses $m_{X_1}$ and $m_{X_2}$, respectively.
We then have $m_{X_1} + m_{X_2} \approx 4$~GeV, from the preceding discussion.
For $y_{ij}\sim 1$, one could easily have a mass eigenstate that is
lighter than a nucleon.  For example, with $m_1 = 2.2$~GeV, $m_2 = 1.8$~GeV, and $m_{12}=m_{21} = 1.4$~GeV,
we find $m_{X_1}=0.6$~GeV and $m_{X_2}= 3.4$~GeV.
Hence, one of the ADM fields, hereafter denoted by $X_1$,
could be generically lighter than a nucleon without tuning of model parameters.

The above two-component ADM model then turns out to have an interesting signature, namely
the decay of protons and neutrons into DM!  Such {\it dark nucleon decay} (DND)
processes are allowed for $m_{X_1}\lsim m_N-m_\pi$, where $m_N$
is the nucleon mass and $m_\pi$ is the mass of the pion.  Note that within the inflationary reheating
scenario ({\it i.e.} with $w=0$) discussed above, we end up with scales $\Lambda_{1,2} \gsim 10^{16}$~GeV,
close to typical Grand Unified Theory (GUT) scales. This suggests that the associated
DND rates would be in a range accessible to current or future experiments \cite{NDEXP}.
These estimates were obtained assuming $M_c \sim \rmP$.  However,
(quantum) gravity effects at very high scales could
yield a range of values for $M_c$.  Hence, one may assume
$M_c = \kappa  \rmP$, with $\kappa \sim 0.05-1$, at the level of our effective theory treatment.
In that case, Eqs.~(\ref{OBX}) and (\ref{nBs;w=0}) imply
nucleon lifetimes of order $10^{34-36}$~yr can result from dark decays.

The DND discussed above is not entirely ``dark" as it
includes a meson: $N \to X_1 \pi, X_1 K, \ldots$.  Such processes
will mimic standard nucleon decay (SND) into a neutrino and a meson, $N\to \nu \pi, \nu K, \ldots$,
predicted in other contexts, like GUT scenarios.  However, for DND the meson kinematics
are expected to be different from that of the SND  involving a neutrino.  (This is reminiscent of
annihilation of nucleons in scattering from ADM, as discussed in
Refs.\cite{hylogenesis,IND}.)  The meson momenta for $N\to \nu \pi (K)$ and
$N\to X_1 \pi (K)$ could differ significantly, depending on the value of $m_{X_1}$.  As expected, the
lighter the DM particle $X_1$, the more similar DND and SND are.  For instance,
the SND meson momenta are given by
$p^{\rm SND}_{\pi(K)} \approx 460 (340)$~MeV (see Table \ref{momenta}), whereas with
$m_{X_1}=300$~MeV we have $p^{\rm DND}_{\pi(K)} \approx 410 (245)$~MeV.  However, if $m_{X_1} = 600$~MeV,
we find $p^{\rm DND}_{\pi} \approx 250$~MeV and the kaon mode is forbidden.  Thus, for somewhat
heavier $X_1$, we find markedly different pion kinematics.

As discussed in Ref.\cite{IND}, kinematic differences can help distinguish novel decay channels such as DND.
On the other hand, applying the existing nucleon lifetime constraints to the new physics must be done with care,
as kinematics can affect the detection efficiency.  For example, if $m_{X_1}$ is sufficiently close
to the kinematic limit for allowed decays, the pion can be very slow and below \v{C}erenkov threshold.
On the other hand, in case of neutral meson decays into diphotons, the suppressed boost could help
in resolving the photon pair and event reconstruction.

\begin{table}
\begin{tabular}{|c|c|c|}
\hline
Decay Channel & $p_{\rm meson}^{\rm SND}$ (MeV) &
$(p_{\rm meson}^{\rm DND}, m_{X_1})$ (MeV) \\
\hline
$N \to \pi$ & 460 & (410, 300)\\
~ & ~ & (250, 600)\\
\hline
$N \to K$ & 340 &  (245, 300) \\
~ & ~ & ($-$, 600) \\
\hline
\end{tabular}
\caption{Sample meson momenta in standard and
dark nucleon decays, for $m_{X_1} = 300, 600$~MeV.}
\label{momenta}
\end{table}

Here, we would like to add that there are other cosmological epochs governed by $w \neq 0$
in which one could consider the interaction (\ref{OQ}) \cite{GB}.  For example,
using the results of Ref.\cite{GB}, it seems that $w\sim 3/4$ and $\Lambda \sim {\rm few}\times 10^{12}$~GeV could yield
$n_B/s \sim 10^{-10}$.  For values of $\Lambda$ in this range, $\tau_X\sim 10^{27}$~s is near the current
observational bounds \cite{Bell:2010fk} for $m_X\sim 2$~GeV (DND irrelevant)
and could lead to an indirect DM decay signal.
However, the post-inflationary reheating characterized
by $w = 0$, considered above, is well-motivated and can be a generic feature of
standard cosmological scenarios.  We expect that our main conclusions
can be accommodated by various conventional inflationary models.  We also note that
since sufficient asymmetry for $w=0$ requires
inflationary scales $V_I^{1/4} \gsim 10^{16}$~GeV, this scenario would typically
suggest that the detection of tensor mode perturbations could be within observational
reach \cite{Planck}.

Before closing, we will examine a variant scenario that includes
a light vector boson $Z_d$ associated with a dark $U(1)_d$ force that mediates $X\bar X$
annihilation \cite{Pospelov:2007mp}.  Here, $Z_d$ couples
to $X$ and kinetically mixes with the photon; such a vector is often
referred to as a ``dark photon"\cite{darkZ}.  This setup avoids potential
problems with a long-lived $S$ when $m_X<1$~GeV, as in the interesting
case with multiple ADM fermions.
For operators of the type (\ref{OBX}) to be allowed, we have to assume that $X_R$ is not charged under
$U(1)_d$.  However, $X_L$ can have the required dark gauge charge.  To allow
a Yukawa coupling of the type $y_X S \bar X_L X_R$, the
scalar $S$ needs to be charged under the dark force.  This will also ensure that
$Z_d$ will have a non-zero mass $m_{Z_d}\sim g_d \vev{S}$, where $g_d$ is the $U(1)_d$
gauge coupling.  For $m_{Z_d} \ll m_X $, the annihilation process
$X \bar X \to Z_d Z_d$ has a cross section  
\beq
\sigma(X \bar X\to Z_d Z_d) v\sim \frac{g_d^2\, y_X^2}{32 \pi m_X^2}, 
\label{XtoZd}
\eeq
where $v$ is relative velocity.      
For instance, with $g_d \sim y_X/10 \sim 0.1$ 
and $m_X\sim 1$~GeV, we find an annihilation cross section
$\sim 40$~nb which is quite sufficient for removing the symmetric $X$ density.

The lifetime of $Z_d$ depends on the degree of kinetic mixing
with the photon parameterized by $\eps$.  For $m_{Z_d} \sim 100$~MeV,
experimental bounds require $\eps\lsim {\rm few}\times 10^{-3}$ (for a summary of recent
constraints see, {\it e.g.}, Refs.~\cite{Davoudiasl:2013aya,Andreas:2013iba}).
However, to avoid conflict with direct detection bounds for $m_{X_2}\sim {\rm few}$~GeV, we may need to consider
$\eps \lsim 10^{-5}$ \cite{hylogenesis}.
For these values, $Z_d$ lifetime is of order
$16 \pi/(\eps^2 \,m_{Z_d})\gsim 10^{-11}$~s which does not pose a difficulty.
Here, we have assumed that the decays of $Z_d$ are dominated by the visible $e^+e^-$ channel.  This is
consistent with our requirement $m_X>m_{Z_d}$ (see
Refs.\cite{Batell:2009di,deNiverville:2011it,Izaguirre:2013uxa}
for a discussion of $Z_d$ phenomenology when dark matter final states are dominant).

In summary, we extended ``gravitational baryogenesis" to accommodate
the generation of a dark matter asymmetry.  We focused on the well-motivated
matter dominated cosmological equation of state ($w=0$) that characterizes reheating
through inflaton oscillations.  The minimal required extension can be implemented
by adding a singlet Dirac fermion dark matter and a singlet scalar.  This setup leads to
an unstable dark matter with a lifetime of order $10^{34}$~yr.  The minimal model (for $w=0$)
does not offer readily accessible observational signatures.  However, a modestly enlarged dark sector
with more than one fermion could lead to the interesting possibility of nucleon decays into
dark matter fermions.  We showed, as an example, that a simple two-fermion
dark matter sector could lead to dark nucleon decays, with inverse rates $10^{34-36}$~yr, 
relevant to current or future experiments.
In our reference $w=0$ cosmology, the large inflationary scale required for sufficient
dark matter and baryon asymmetries typically suggests that tensor mode perturbations may be within
reach of astrophysical measurements.

\acknowledgments

We thank R. Kitano for collaboration during the early stages of this work.
This work is supported in part by the United States Department of Energy
under Grant Contracts DE-AC02-98CH10886.



\end{document}